\documentclass[11pt,twoside,usenatbib]{article}
\usepackage{asp2010}

\resetcounters

\begin{document}

\vspace{-3mm}
\title{Why variable AGB stars with Long Secondary Periods aren't binaries, but are dusty}
\vspace{-3mm}
\author{Christine Nicholls$^1$
\affil{$^1$Research School of Astronomy \& Astrophysics, Mt. Stromlo Observatory, Cotter Road, Weston Creek ACT 2611 AUSTRALIA}}

\begin{abstract}
Roughly 30\% of variable AGB stars show a Long Secondary Period, or LSP. These LSPs have posed something of a problem in recent years and their cause remains a mystery. By combining VLT-derived velocity curves with MACHO and OGLE light curves we were able to examine many properties of these stars and test the theory that LSPs are caused by binarity. We show why we concluded that the binary model for LSPs is unlikely. Examining mid-infrared SAGE observations for stars with LSPs shows that these stars are surrounded by a significant amount of cool dust in a non-spherical distribution, e.g. a disk or clumps. The unlikeliness of binarity in these stars forces us to conclude that the dust is not in a disk. We are left without an acceptable explanation for Long Secondary Periods in AGB stars.
\end{abstract}

\vspace{-10mm}
\section{Long Secondary Periods: an Enduring Mystery}
Thirty percent of variable red giants have light curves showing a primary oscillation and a Long Secondary Period (LSP). Despite several studies in the past decade, no model has been proposed that can satisfactorily explain the LSPs. Stars exhibiting LSPs lie on a period-luminosity sequence \nocite{wood99mn} (Wood et al.\ 1999).

Proposed explanations for LSPs include motion due to a binary or planetary companion, radial and nonradial pulsation, dust obscuration and chromospheric effects. We examine here the most popular model, that of a binary companion. LSP variables have many properties that could be explained by binarity. These include an eclipse-like lightcurve, periodic radial velocity variations, periods similar to known close red giant binary systems with solar mass components, and small variation in $T_{\rm{eff}}$.

\vspace{-5mm}
\section{Some Dusty Evidence}
Recently, we discovered that LSP variables show a significant mid-IR colour excess compared to variables without LSPs \citep{dust}. This can be seen in fig. 3 of that paper. This colour excess almost certainly indicates the presence of dust. Therefore dust is either a cause, or a product of, the LSP variation. 

Stars with roughly spherical dusty winds show a well-correlated increase in both near-IR (e.g. J-K) colour and mid-IR (e.g. K-[24]) colour with increasing mass loss rate.  However, objects such as RCB stars or RV Tauri stars with patchy or disk-like circumstellar dust show a large mid-IR excess but a variable near-IR colour depending on whether the circumstellar dust lies between the stellar photosphere and the observer.  As shown in fig. 3 of \cite{dust}, the LSP variables lie in the same region of the (K-[24], J-K) diagram as the RCB stars and RV Tauri stars, indicating that the dust around them is patchy or disk-like.

Dusty disks are usually associated with binaries. Therefore, this appears to be evidence in favour of a binary cause for Long Secondary Periods.

\vspace{-4mm}
\section{Some Problems}
However, the binary model has its problems. The LSP velocity curves have a characteristic shape which, for a binary system, suggests an eccentric orbit and a large angle of periastron, $\omega$, with a non-uniform distribution \citep[see fig. 4 of ][]{seqDpaper}.

One expects $\omega$ to be uniformly distributed for binaries. Applying a K-S test, we find the probability that our distribution of $\omega$ is consistent with the uniform distribution is $1.4 \times 10^{-3}$. In other words, the probability that LSP variables are binaries is extremely small. 

Our stars also have low, relatively similar velocity amplitudes, something unexpected in a sample of binaries. The median LSP velocity amplitude in our sample is $3.5\,\rm{km\,s^{-1}}$, and deviation from this value is small. In a binary system, the velocity amplitude and period give an estimate of the mass of the companion. For a typical LSP variable with LSP = 500 days, velocity amplitude = $3.5\,\rm{km\,s^{-1}}$, and assuming $M_{\rm{tot}} = 1.5\,M_{\odot}$, the orbital separation is $\sim$1.4 AU and the companion has a mass of 0.09$\,M_{\odot}$.

This characteristic companion mass is problematic: there is an observed deficit of binary companions of $\sim\,0.09\,M_{\odot}$ to main sequence stars, compared to both more massive stellar companions and less massive planetary companions. This is known as the `Brown Dwarf Desert'. Indeed, using the results of \cite{browndwarfdesert} we calculate only 0.86\% of solar-vicinity main sequence stars should have companions between 0.06 and 0.12$\,M_{\odot}$. Given that $\sim\,30\%$ of AGB stars show LSPs, if LSP variables are binaries we would expect $\sim\,30\%$ of stars to have companions in this mass range.

Some suggest that LSPs may be caused by ellipsoidal variation, where a red giant is distorted by a close companion. The light curves of ellipsoidal variables should complete two cycles for every velocity curve cycle, as the light variation is dominated by the star's ellipsoidal shape, while the velocity curve is dominated by orbital motion. However we recently showed \citep{seqEpaper} that LSP variables do not show ellipsoidal variations. 

\vspace{-4mm}
\section{Does Dust Hold the Answer?}
LSP variables show a significant mid-IR excess that indicates dust in a non-spherical distribution. Although this normally indicates a dusty disk, these are usually associated with binaries, and we have presented convincing evidence against a binary model for LSPs above. At this stage, we can say that LSP variables probably have asymmetrical dust ejections, but cannot shed any further light on the underlying cause of LSPs.

\vspace{-4mm}

\bibliographystyle{asp2010}
\bibliography{bibliographynew}

\begin{thebibliography}{}
\expandafter\ifx\csname natexlab\endcsname\relax\def\natexlab#1{#1}\fi
\expandafter\ifx\csname url\endcsname\relax
  \def\url#1{\texttt{#1}}\fi
\expandafter\ifx\csname urlprefix\endcsname\relax\def\urlprefix{URL }\fi
\providecommand{\eprint}[2][]{\url{#2}}

\bibitem[{{Grether} \& {Lineweaver}(2006)}]{browndwarfdesert}
{Grether}, D., \& {Lineweaver}, C.~H. 2006, \apj, 640, 1051.
  \eprint{arXiv:astro-ph/0412356}

\bibitem[{{Nicholls} et~al.(2010){Nicholls}, {Wood}, \& {Cioni}}]{seqEpaper}
{Nicholls}, C.~P., {Wood}, P.~R., \& {Cioni}, M. 2010, \mnras, 405, 1770.
  \eprint{1002.3651}

\bibitem[{{Nicholls} et~al.(2009){Nicholls}, {Wood}, {Cioni}, \&
  {Soszy{\'n}ski}}]{seqDpaper}
{Nicholls}, C.~P., {Wood}, P.~R., {Cioni}, M., \& {Soszy{\'n}ski}, I. 2009,
  \mnras, 399, 2063. \eprint{0907.2975}

\bibitem[{{Wood} \& {et al. (MACHO Collaboration)}(1999)}]{wood99mn}
{Wood}, P.~R., \& {et al. (MACHO Collaboration)} 1999, in IAU Symp. 191:
  Asymptotic Giant Branch Stars, edited by T.~{Le Bertre}, A.~{Lebre}, \&
  C.~{Waelkens} (San Francisco: Astronomical Society of the Pacific), 151

\bibitem[{{Wood} \& {Nicholls}(2009)}]{dust}
{Wood}, P.~R., \& {Nicholls}, C.~P. 2009, \apj, 707, 573. \eprint{0910.4418}

\end{thebibliography}

\end{document}